\documentclass[showpacs,preprintnumbers,amsmath,amssymb]{revtex4}

\usepackage{graphics}
\usepackage{dcolumn}
\usepackage{bm}
\usepackage{amsmath,epsfig}
\usepackage{tabularx}
\usepackage{latexsym}

\newcommand{\ket}[1]{\left\vert#1\right\rangle}
\newcommand{\bra}[1]{\left\langle#1\right\vert}

\begin{document}

\title{Atomic teleportation via cavity QED and position measurements: efficiency analysis}
\author{Michele Tumminello$^1$ and Francesco Ciccarello$^{1,2}$}

\affiliation{%
$^1\,$Dipartimento di Fisica e Tecnologie Relative,
Universit\`a di Palermo, Viale delle Scienze, Edificio 18, I-90128
Palermo, Italy.\\
$^2\,$Consorzio Nazionale Interuniversitario per le
Scienze Fisiche della Materia (CNISM), Italy, NEST-INFM (CNR)
\& Dipartimento di Scienze Fisiche ed Astronomiche, Universit\`{a}
degli Studi di Palermo, Via Archirafi 36, I-90123 Palermo, Italy.
}%

\begin{abstract} 
We have recently presented a novel protocol to teleport an
unknown atomic state via cavity QED and position measurements. Here,
after a brief review of our scheme, we provide a quantitative study
of its efficiency. This is accomplished by an explicit description
of the measurement process that allows us to derive the fidelity
with respect to the atomic internal state to be teleported.
\end{abstract}


\maketitle
\section{Introduction}
\label{intro} The key role played by quantum entanglement in a
number of crucial quantum information processing tasks is now firmly
grounded \cite{nc}. One of the most striking applications harnessing
such powerful resource is the teleportation of an unknown qubit, the
unit of quantum information, between two distant systems first
proposed in famous paper by Bennett \emph{et al.} \cite{bennett}.
The idea is essentially to transfer an unknown quantum state from an
input qubit $A$ to a target qubit $B$ by using an \emph{ancilla},
e.g. a third auxiliary qubit $C$. Teleportation is achieved via
preparation of initial maximally entangled states between qubits $B$
and $C$ and successive Bell measurements on $A$ and $C$. This
phenomenon soon obtained large attention especially after in its
experimental demonstrations \cite{bowm-boschi,NMR,natphys,nature}.

Of course, due to its purely quantum nature, teleportation can be
attained in systems that exhibit a fully quantum behaviour. This is
one of the reasons why cavity quantum electrodynamics (QED), where
coherent coupling between Rydberg atoms and the quantized
electromagnetic (e.m.) field is possible \cite{raimond}, has been
soon addressed as a promising scenario to achieve teleportation
\cite{teleportation-QED}. In cavity-QED schemes for teleporting
states between two atoms, the typical strategy is indeed to harness
the coherent atom-field interaction using cavity e.m. modes and/or
additional atoms as ancillary systems.

Among these works are some recent schemes where direct (in general
quite difficult) projections onto Bell states are avoided
\cite{vaidman,de-almeida,zheng,ye-guo,cardoso}. In particular, Zheng
has proposed a scheme for approximately teleporting an unknown
internal state between two atoms which successively interact with a
cavity mode according to the Jaynes-Cummings Hamiltonian
\cite{zheng}. Ye and Guo have presented another scheme that does not
require projections onto Bell-state and makes use of three atoms and
a single-mode cavity field out of resonance \cite{ye-guo}. The
atom-atom coupling via the virtual excitations of the cavity field
is exploited for teleporting a quantum state between two atoms. The
probability of success of the Zheng scheme is 1/4, whereas the
teleportation is successful with probability 1/2 in the proposal by
Ye and Guo.  Noticeably, both the schemes require precise tuning of
the atom-cavity field interaction time \cite{zheng,ye-guo}.

Very recently, we have proposed the first cavity-QED scheme that
exploits the atomic translational dynamics in order to accomplish
teleportation between two atoms \cite{TC}. Among the major
advantages are the probability of success of 1/2, the need to
measure only product states as well as the ability of the protocol
to work without any holonomous constraint on the atom-photon
interaction times. In particular, the latter feature implies that,
unlike other schemes \cite{zheng,ye-guo} no precise tuning of the
atom-cavity field interaction times is required. It only suffices
that such times are large enough in order for \emph{which-path}
information of the two atoms to become accessible.

In Ref. \cite{TC}, we have provided evidence that position
measurements enable successful teleportation in the regime of
accessible which-path information. This phenomenon was shown to stem
from the asymptotic orthogonality of the field-deflected atomic
wavepackets \cite{Vag-Cus,which-path}. In this work, we make such
conclusions more exact by explicitly including the measurement
process of the atomic positions. This allows us to calculate the
fidelity with respect to the state to be teleported as a function of
the parameters entering the dynamics. We derive a lower bound for
such a quantity and study its dependence on the measured atomic
positions and the atom-cavity interaction times.

This paper is organized as follows. In Sec. \ref{syst-appr}, we
introduce the system and the Hamiltonian. In Sec. \ref{QT-section},
we review the teleportation scheme of Ref. \cite{TC}. In Sec.
\ref{meas-process}, we calculate the final state of the atom-field
system once all the necessary measurements required for successful
teleportation have been explicitly taken into account. In Sec.
\ref{fidelity} we analyze the efficiency of the scheme in terms of
the fidelity with respect to the atomic state to be teleported. We
also provide a lower bound for such a quantity and investigate its
behaviour as a function of the measured atomic positions and the
atom-field interaction times. Finally, in Sec. \ref{conclusions} we
draw our conclusions.

\section{System and approach}
\label{syst-appr}

We consider two identical two-level atoms, labeled 1 and 2, of mass
$m$ and Bohr frequency $\omega$. The atoms interact in succession
with the e.m. field of the same e.m. cavity. We assume that the
velocity of each atom along the $z$-direction (orthogonal to the
$x$-cavity axis) is large enough that the motion along the $z$-axis
is not affected by the cavity field and can be treated classically.
Denoting by $a$ and $a^{\dag}$ the annihilation and creation
operators of the cavity field and assuming the resonance condition,
the free Hamiltonian $H_0$ can be written as
\begin{equation}\label{H0}
H_0=\sum_{i=1,2}\left[\frac{\hat{p}_i^2}{2m}+\hbar \omega
S_{z,i}\right]+\hbar \omega a^{\dag}a \,,
\end{equation}
where -- for each atom $i=1,2$ -- $S_{z,i},S_{\pm,i}$ are the usual
spin-1/2 operators and $\hat{p}_i=-i\hbar(d/dx_{i})$ is the
$x$-component of the momentum operator. In the Rotating Wave
Approximation, each atom $i$ couples to the cavity field according
to the interaction Hamiltonian
\begin{equation}\label{Hif}
H_{if}=\hbar \varepsilon \sin (k\hat{x}_{i})
\left(a^{\dag}S_{-,i}+aS_{+,i}\right) \,\,\,\,(i=1,2)
\end{equation}
with $k$ and $\varepsilon$ standing for the wave number of the e.m.
mode and the atom-field coupling constant, respectively, and where
$\hat{x}_{i}$ is the $i$th atomic position operator along the cavity
axis.

Hamiltonian (\ref{Hif}) accounts for the spatial structure of the
e.m. field along the $x$-cavity axis. Based on such Hamiltonian
model, a number of observable phenomena have been shown such as the
optical Stern-Gerlach effect \cite{SGE}, self-induced transparency
\cite{schlicher}, modulation of the atomic decay in a damped cavity
\cite{wilkens} and non-dissipative damping of the Rabi oscillations
\cite{Vag-Cus,which-path}.

When both the atomic wavepackets have width $\sigma_{x_{i}}$ small
enough compared with the cavity wavelength $2\pi/k$
($\sigma_{x_{i}}\ll 2\pi/k$) and in a nodal region of the cavity
$H_{i}$, can be approximated as \cite{nota-nodale}
\begin{eqnarray}\label{Hif-nodal}
H_{iN}&=&\hbar \varepsilon k\, \hat{x}_{i}
\left(a^{\dag}S_{-,i}+aS_{+,i}\right),
\end{eqnarray}
where $\hat{x}_{i}$ stands for the atomic position operator of the
$i$th atom with respect to the nodal point.

At time $t=0$, atom 1 enters the cavity and interacts with the field
for a time $t_{1}$. At a later time $t_{2}>t_{1}$, atom 2 enters the
cavity and couples to the field state modified by the first atom. At
time $t_{3}>t_{2}$ atom 2 exits the cavity. At times $t\geq t_{3}$
both the atoms are therefore out of the cavity and evolve freely. In
the interaction picture, the Hamiltonian at all times reads
\begin{eqnarray}\label{H-nodal}
  H_{N}^{I}(t)&=&\hbar\varepsilon k\left(\hat{x}_{1}+\frac{\hat{p}_{1}}
  {m}t\right)\mu_{t}(0,t_{1})u_{1}+ \hbar\varepsilon k\left(\hat{x}_{2}+\frac{\hat{p}_{2}}{m}t\right)
  \mu_{t}(t_{2},t_{3})u_{2} \,,
\end{eqnarray}
where we have introduced the atom-field operators
$u_{i}=a^{\dag}S_{-,i}+aS_{+,i}$ and where the time interval during
which each atom interacts with the cavity mode is accounted for by
means of the function $\mu_{t}(t',t'')=\theta(t-t')-\theta(t-t'')$,
$\theta(t)$ being the usual Heaviside function. Of course, in the
time interval $[t_1,t_2]$ and for $t\geq t_3$  $H_{N}^{I}(t)$
vanishes since no atom is inside the cavity. The Hamiltonian
operator of Eq. (\ref{H-nodal}) can be used to derive the exact
dynamics of a given initial state of the two-atom-field system at
times $t\geq t_3$ through the respective evolution operator
$U_{N}^{I}(t\geq t_{3})$
\begin{equation}\label{Ualpha}
U_{N}^{I}(t\geq t_{3})=T\,\exp\left[-\frac{i}{\hbar}\int_{0}^{t_3}
H_{N}^{I}(t)dt\right]
\end{equation}
with $T$ standing for the time-ordering operator and where the
second integration bound is due to the fact that $H_{N}^{I}=0$ for
$t\geq t_{3}$.

Due to the fact that atom 2 enters the cavity after atom 1 has come
out of it, it is possible to split up $U_{N}^{I}(t\geq t_{3})$ into
the product of two evolution operators $U_{N,1}^{I}(t\geq t_{3})$
and $U_{N,2}^{I}(t\geq t_{3})$. Each operator $U_{N,i}^{I}(t\geq
t_{3})$ only affects the dynamics of atom $i$. In formulae (from now
on, whenever unnecessary, the time argument ``$(t\geq t_{3})$"
and/or the apex ``$I$"  in the evolution operators will be omitted)
\begin{equation}\label{U-alpha}
U_{N}=U_{N,2} \cdot U_{N,1}
\end{equation}
with
\begin{eqnarray}
U_{N,1}=T\,\exp\left[-\frac{i}{\hbar}\int_{0}^{t_1} H_{N}^{I}(t)dt\right]=U_{N,1}(\hat{x}_1,\hat{p}_1,u_1), \label{Ualpha1}\\
U_{N,2}=T\,\exp\left[-\frac{i}{\hbar}\int_{t_2}^{t_3}
H_{N}^{I}(t)dt\right]=U_{N,2}(\hat{x}_2,\hat{p}_2,u_2),\label{Ualpha2}
\end{eqnarray}
where in the right-hand side of both equations we have explicitly
indicated the quantities the $U_{N,i}$'s depends on according to Eq.
(\ref{H-nodal}).

\section{Teleportation scheme} \label{QT-section}

We denote the ground and excited states of the $i$th atom by
$\ket{g_{i}}$ and $\ket{e_{i}}$, respectively. Assume that atom 2 is
the one whose initial internal state, say $\ket{\alpha}_{2}$, is to
be teleported. Such state is written as
\begin{equation}\label{state-teleported}
\ket{\alpha}_{2}=\cos
\frac{\vartheta}{2}\ket{e_2}+e^{i\varphi}\sin\frac{\vartheta}{2}\ket{g_2}
\end{equation}
with $\vartheta\in[0,\pi]$ and $\varphi\in[0,\pi]$.

By indicating the Fock states of the cavity field as $\ket{n}$
($n=0,1,...$), we consider the initial state of the system
$\ket{\Psi(0)}=\ket{\varphi_1(0)}\ket{e_1}\,\,
\ket{\varphi_2(0)}\ket{\alpha}_{2}\,\,\ket{0}$ where
$\ket{\varphi_i(0)}$ (associated with each atom $i=1,2$) is a
Gaussian wavepacket of minimum uncertainty, such that the product
between the initial position and momentum widths fulfills
$\sigma_{x_{i}}\cdot\sigma_{ p_{i}}= \hbar/2$. Consider now the
usual dressed states of the $i$th atom
$\ket{\chi_{n,i}^{\pm}}=\left(\ket{e_{i}}\ket{n}\pm
\ket{g_{i}}\ket{n+1}\right)/\sqrt{2}$ ($n=0,1,...$). These states
are eigenstates of the $u_i$ operators since
$u_i\ket{\chi_{n,i}^{\pm}}=\pm \sqrt{n+1} \ket{\chi_{n,i}^{\pm}}$
(while $u_{i}\ket{g_{i}}\ket{0}=0$). The dressed states together
with $\ket{g_{i}}\ket{0}$ ($i=1,2$) represent an orthonormal basis
of the corresponding Hilbert space. As $u_i$ commutes with $U_{N,i}$
according to Eqs. (\ref{H-nodal}), (\ref{Ualpha1}) and
(\ref{Ualpha2}), the effective representation $U_{N,i}^{(n,\pm)}$ of
$U_{N,i}$, as applied to a dressed state $\ket{\chi_{n,i}^{\pm}}$,
is obtained by simply replacing $u_i$ with $\pm \sqrt{n+1}$ in Eqs.
(\ref{Ualpha1}) and (\ref{Ualpha2}). This yields
\begin{equation}\label{U-Ni-eff}
U_{N,i}^{(n,\pm)}=U_{N,i}(\hat{x}_i,\hat{p}_i,\pm
\sqrt{n+1})\,\,\,\,\,\,\,(n=0,1,...),
\end{equation}
while the effective representation of $U_{N,i}$ -- as applied to
state $\ket{g_{i}}\ket{0}$ -- reduces to the identity operator for
both the atoms $i=1,2$. The operators in Eq.~(\ref{U-Ni-eff})
clearly affects only the atomic translational dynamics and therefore
allows to define a family of atomic translational wavepackets
$\ket{\Phi_{n,i}^{\pm}}$ according to
$\ket{\Phi_{n,i}^{\pm}}=U_{N,i}^{(n,\pm)}\ket{\varphi_i(0)}$ such
that
\begin{equation} \label{stati_phi2}
U_{N,i} \ket{\varphi_i(0)}\ket{\chi_{n,i}^{\pm}}=\ket{
\Phi_{n,i}^{\pm}}\ket{\chi_{n,i}^{\pm}}.
\end{equation}
Once the time evolution operator (\ref{U-alpha}) is applied to the
initial state $\ket{\Psi(0)}$, the state of the whole system at a
time $t \ge t_{3}$ -- when both the atoms are out of the cavity --
can be written in the form
\begin{eqnarray}\label{expansion-t3}
\ket{\psi(t_3)}=\ket{\lambda_{0,1}}\ket{\varphi_2(0)}\ket{g_2}\ket{0}\nonumber
+\sum_{n=0,1}\sum_{\eta=-,+}\left(\ket{\lambda_{n,1}^{\eta}}\ket{\Phi_{n,2}^{\eta}}\ket{\chi_{n,2}^{\eta}}\right),
\end{eqnarray}
where the $\lambda$ states of atom 1 are defined according to
\begin{eqnarray}\label{lambda}
\ket{\lambda_{0,1}}&=&\left(\frac{\ket{\Phi_{0,1}^{+}}+\ket{\Phi_{0,1}^{-}}}{2}\right)\,e^{i\varphi} \sin\frac{\vartheta}{2}\ket{e_1},\\
\ket{\lambda_{0,1}^{\pm}}&=&\left(\frac{\ket{\Phi_{0,1}^{+}}+\ket{\Phi_{0,1}^{-}}}{2\sqrt{2}}\right)\,\cos\frac{\vartheta}{2}\ket{e_1}
\pm  \left(\frac{\ket{\Phi_{0,1}^{+}}-\ket{\Phi_{0,1}^{-}}}{2\sqrt{2}}\right) \,e^{i\varphi}\sin\frac{\vartheta}{2}\ket{g_1}, \\
\ket{\lambda_{1,1}^{\pm}}&=&\left(\frac{\ket{\Phi_{0,1}^{+}}-\ket{\Phi_{0,1}^{-}}}{2\sqrt{2}}\right)\cos\frac{\vartheta}{2}\ket{g_1}.
\end{eqnarray}
By indicating the time spent inside the cavity by atoms 1 and 2 with
$\tau_1=t_2-t_1$ and $\tau_2=t_3-t_2$, respectively, the states
$\ket{\Phi_{n,i}^{\pm}}$ appearing in Eq.~(\ref{expansion-t3})
fulfill the following important property
\cite{which-path,epl-2atoms,epjd}
\begin{eqnarray}\label{cond-stati-phi}
\lim_{\tau_i\rightarrow
\infty}\bra{\Phi_{n,i}^{+}}\Phi_{n,i}^{-}\rangle =0. \label{prop2}
\end{eqnarray}
According to Eq.~(\ref{cond-stati-phi}), wavepackets
$\ket{\Phi_{n,i}^{+}}$ and $\ket{\Phi_{n,i}^{-}}$ exhibit a
negligible overlap for long enough times of flight $\tau_i$. Times
of flight of the order of a few Rabi oscillations are sufficient in
order to get negligible overlapping \cite{epl-2atoms,epjd}. Such
noticeable circumstance allows to distinguish the elements of the
set of translational states \{$\ket{\Phi_{n,i}^{\pm}}$\} through
measurements of the atomic positions along the $x$-axis. As can be
shown, Eq.~(\ref{cond-stati-phi}) yields that all the terms
appearing in (\ref{expansion-t3}) are mutually orthogonal provided
$\tau_1$ and $\tau_2$ are sufficiently large. By expressing the
dressed states $\ket{\chi_{n,2}^{\pm}}$ appearing in
Eq.~(\ref{expansion-t3}) in terms of states $\ket{g_2}\ket{n}$ and
$\ket{e_2}\ket{n}$, one recognizes the occurrence of cases where
measurements of the photon number, of the internal state of atom 2
and of the positions of the two atoms can make atom 1 collapse into
the initial internal state of atom 2 ($\ket{\alpha}_2$), \emph{i.e.}
a successful teleportation can take place.
\begin{table}
\begin{tabular}{||c|c|c|c|c|c|c||}
\cline{1-7}
$\mathbf{\ket{n}}$ & \bf{Int. 2} & \bf{Tr. 1} & \bf{Tr. 2} & \bf{Result} & \bf{Int. 1} & $\mathbf{P_{fail}}$ \\
\cline{1-7}
2 & -- & -- & -- & Unsuccessful & -- &  $\frac{1}{8}(1+ \cos \vartheta)$\\
\cline{1-7}
$\,$ & $\ket{e_2}$ & -- & -- & Unsuccessful & -- & $\frac{1}{8}(1+ \cos \vartheta)$ \\
\cline{2-7}
$\,$ & $\ket{g_2}$ & $\ket{\Phi_{0,1}^{-}}$ & $\ket{\Phi_{0,2}^{-}}$ & Successful & $\cos \frac{\vartheta}{2}  \ket{e_1} +e^{i \varphi} \sin \frac{\vartheta}{2}  \ket{g_1} $ & -- \\
\cline{2-7}
1 & $\ket{g_2}$ & $\ket{\Phi_{0,1}^{-}}$ & $\ket{\Phi_{0,2}^{+}}$ & Successful \cite{footnote} & $\cos \frac{\vartheta}{2}  \ket{e_1} -e^{i \varphi} \sin \frac{\vartheta}{2}  \ket{g_1} $ & -- \\
\cline{2-7}
$\,$ & $\ket{g_2}$ & $\ket{\Phi_{0,1}^{+}}$ & $\ket{\Phi_{0,2}^{+}}$ & Successful & $\cos \frac{\vartheta}{2}  \ket{e_1} +e^{i \varphi} \sin \frac{\vartheta}{2}  \ket{g_1} $ & -- \\
\cline{2-7}
$\,$ & $\ket{g_2}$ & $\ket{\Phi_{0,1}^{+}}$ & $\ket{\Phi_{0,2}^{-}}$ & Successful \cite{footnote} & $\cos \frac{\vartheta}{2}  \ket{e_1} -e^{i \varphi} \sin \frac{\vartheta}{2}  \ket{g_1}$ & -- \\
\cline{1-7} \cline{1-7}
$\,$ & $\ket{g_2}$ & -- & -- & Unsuccessful & -- & $\frac{1}{4}(1- \cos \vartheta)$ \\
\cline{2-7}
$\,$ & $\ket{e_2}$ & $\ket{\Phi_{0,1}^{-}}$ & $\ket{\Phi_{0,2}^{-}}$ & Successful & $\cos \frac{\vartheta}{2}  \ket{e_1} +e^{i \varphi} \sin \frac{\vartheta}{2}  \ket{g_1} $ & -- \\
\cline{2-7}
0 & $\ket{e_2}$ & $\ket{\Phi_{0,1}^{-}}$ & $\ket{\Phi_{0,2}^{+}}$ & Successful \cite{footnote} & $\cos \frac{\vartheta}{2}  \ket{e_1} -e^{i \varphi} \sin \frac{\vartheta}{2}  \ket{g_1}$ & -- \\
\cline{2-7}
$\,$ & $\ket{e_2}$ & $\ket{\Phi_{0,1}^{+}}$ & $\ket{\Phi_{0,2}^{+}}$ & Successful & $\cos \frac{\vartheta}{2}  \ket{e_1} +e^{i \varphi} \sin \frac{\vartheta}{2}  \ket{g_1} $ & -- \\
\cline{2-7}
$\,$ & $\ket{e_2}$ & $\ket{\Phi_{0,1}^{+}}$ & $\ket{\Phi_{0,2}^{-}}$ & Successful \cite{footnote} & $\cos \frac{\vartheta}{2}  \ket{e_1} -e^{i \varphi} \sin \frac{\vartheta}{2}  \ket{g_1}$ & -- \\
\cline{1-7}
\end{tabular}
\bigskip \caption{\label{table}Teleportation measurement scheme. Each case is
represented by given outcomes of the number of photons (1$^{st}$
column), the internal state of atom 2 (2$^{nd}$ column) and the
deflected wavepackets (3$^{th}$ and 4$^{th}$ columns). In the
5$^{th}$ column it is indicated whether or not teleportation has
been successful. If successful, the state onto which atom 1 is
projected, \textit{i.e} $\ket{\alpha}_1$ or
$\ket{\alpha'}_1=-\sigma_z\ket{\alpha}_1$ ($\sigma_z$ is the usual
Pauli matrix), is presented (6$^{th}$ column). If unsuccessful, the
associated unconditional failure probability $P_{fail}$ is given in
the last column.}
\end{table}
For instance, a photon-number measurement signaling a single photon
in the cavity projects $\ket{\psi(t_3)}$ onto the cavity field state
$\ket{1}$. This event occurs with probability $(3+\cos\vartheta)/8$.
Assume now that a further measurement of the internal state of atom
2 is made. If the outcome of such measurement is $\ket{e_2}$, atom 1
is projected onto the ground state $\ket{g_1}$ and thus no
teleportation of the initial state of atom 2 has occurred. The
unconditional probability for this failing event is calculated as
$P_{fail}=(1+\cos\vartheta)/8$. However, it can be noticed that if
atom 2 is found in the ground state $\ket{g_2}$ a further
translational measurement on the two atoms with outcomes
$\ket{\Phi_{0,1}^+}\ket{\Phi_{0,2}^+}$ or
$\ket{\Phi_{0,1}^-}\ket{\Phi_{0,2}^-}$ projects atom 1 onto state
$\ket{\alpha}_{1}=\cos
\frac{\vartheta}{2}\ket{e_1}+e^{i\varphi}\sin\frac{\vartheta}{2}\ket{g_1}$.
This means that the initial internal state of atom 2
($\ket{\alpha}_2$) has been in fact teleported into atom 1. On the
other hand, when the wavepackets
$\ket{\Phi_{0,1}^+}\ket{\Phi_{0,2}^-}$ or
$\ket{\Phi_{0,1}^-}\ket{\Phi_{0,2}^+}$ are found (after that the
state $\ket{g_2}$ has been measured) atom 1 collapses into state
$\cos
\frac{\vartheta}{2}\ket{e_1}-e^{i\varphi}\sin\frac{\vartheta}{2}\ket{g_1}=\ket{\alpha'}_1=-\sigma_z\ket{\alpha}_1$
($\sigma_z$ is the usual Pauli matrix). Clearly, $\ket{\alpha'}_1$
can be straightforwardly transformed into $\ket{\alpha}_1$ through a
$\pi$-rotation around the $z$-axis in order to faithfully reproduce
the initial state of atom 2 and complete the teleportation. In a
similar way, it turns out that when the field vacuum state $\ket{0}$
is found the outcome $\ket{g_2}$ cannot transfer the initial state
of atom 2 into atom 1, whereas successful teleportation is attained
when atom 2 is found to be in the excited state $\ket{e_2}$. All the
possible outcomes of the protocol are summarized in Table
\ref{table}. For each case -- corresponding to given outcomes of the
cavity Fock state $\ket{n}$ (1$^{st}$ column), the internal state of
atom 2 (2$^{nd}$ column), and the two atomic wavepackets (3$^{th}$
and 4$^{th}$ columns) -- it is shown whether or not teleportation
has been successful (5$^{th}$ column). If successful, the state onto
which atom 1 is projected ($\ket{\alpha}_1$ or
$\ket{\alpha'}_1=-\sigma_z\ket{\alpha}_1$) is presented (6$^{th}$
column). If unsuccessful, the associated unconditional failure
probability $P_{fail}$ is given (last column). The total failure
probability, obtained as the sum of the unconditioned failure
probabilities (last column of Table I), is 1/2. Teleportation is
thus successful with probability 1/2. Remarkably, notice how only
\emph{local} measurements on the two atoms and the cavity field are
required. Direct projections onto highly entangled states are thus
avoided. Furthermore, unlike other cavity-QED protocols
\cite{zheng,ye-guo} the interaction time of each atom with the
cavity does not need to fulfill any holonomous constraint. It is
only required that it is large enough in order for
(\ref{cond-stati-phi}) to hold with reasonable approximation.

\section{Measurement process} \label{meas-process}

The discussion developed in the previous Section should make it
clear how the teleportation scheme works. However, the translational
measurements of the atomic wavepackets $\ket{\Phi_{0,i}^+}$ and
$\ket{\Phi_{0,i}^-}$, even though compatible with property
(\ref{cond-stati-phi}), do not formally correspond to position
measurements. In this section, we therefore aim at describing more
explicitly the measurement process required in order to attain
successful teleportation.

To teleport the initial unknown state of atom 2 into atom 1, we need
to perform measurements on the overall Hilbert space of the
ancillary system, \textit{i.e.} the cavity mode and the
translational degrees of freedom of both atoms, and on the internal
degrees of freedom of the input atom 2. Looking at Table \ref{table}
we see that a necessary condition for successful teleportation is
that measurements of the internal state of atom 2 and of the
photon-number respectively give the outcomes $\ket{g_2}$ and
$\ket{1}$ or $\ket{e_2}$ and $\ket{0}$. Setting
$\rho=\ket{\Psi(t_3)}\bra{\Psi(t_3)}$, such measurements project the
system onto state
\begin{equation}\label{int2-campo}
\rho'=\frac{\ket{g_2}\bra{1}\rho(t_3)\ket{g_2}\bra{1}+\ket{e_2}\bra{0}\rho(t_3)\ket{e_2}\bra{0}}{\mathrm{Tr}\left(\ket{g_2}\bra{1}\rho(t_3)\ket{g_2}\bra{1}+\ket{e_2}\bra{0}\rho(t_3)\ket{e_2}\bra{0}\right)}.
\end{equation}
As the denominator equals the probability of successful
teleportation (1/2) in the limit of accessible which-path
information [cfr. Eq. (\ref{cond-stati-phi})] and using Eq.
(\ref{expansion-t3}), the trace of state (\ref{int2-campo}) over the
field and the internal degrees of freedom of atom 2 yields
\begin{eqnarray}
\rho''=\frac{1}{8}\sum_{\mu_1,\mu_2=\pm}\,\sum_{\nu_1,\nu_2=\pm}\left(1+\mu_2\nu_2\right)\ket{\Phi_{0,1}^{\mu_1}}\ket{\Phi_{0,2}^{\mu_2}}\bra{\Phi_{0,1}^{\nu_1}}\bra{\Phi_{0,2}^{\nu_2}}\ket{\mu_1,\mu_2}_1\bra{\nu_1,\nu_2},\label{rho''}
\end{eqnarray}
where
$\ket{\eta,\eta'}_1=\cos\frac{\theta}{2}\ket{e_1}+\eta\eta'e^{i\varphi}\sin\frac{\theta}{2}\ket{g_1}$
($\eta,\eta'=\pm$) is an internal state of atom 1 such that
$\ket{\eta,\eta'}_1=\ket{\alpha}_1$ for $\eta=\eta'$ and
$\ket{\eta,\eta'}_1=\ket{\alpha'}_1=-\sigma_z\ket{\alpha}_1$ for
$\eta\neq\eta'$. Once the first set of measurements have given the
outcomes $\ket{g_2}\ket{1}$ or $\ket{e_2}\ket{0}$ with probability
1/2, assume now to perform further position measurements on the two
atoms along the $x$-cavity axis. If atom 1 and 2 are found at
positions $x_1$ and $x_2$, respectively, the final internal state of
atom 1 $\rho_{1f}$ is obtained by applying the projector
$\ket{x_1,x_2}\bra{x_1,x_2}$ onto (\ref{rho''}) and tracing over the
translational degrees of freedom of both atoms according to
\begin{equation}\label{rho_1f}
\rho_{1f}=\frac{\bra{x_1,x_2}\rho''\ket{x_1,x_2}}{\mathrm{Tr}_1\bra{x_1,x_2}\rho''\ket{x_1,x_2}},
\end{equation}
where $\mathrm{Tr}_1$ stands for the trace over the internal degree
of freedom of atom 1.

\section{Fidelity} \label{fidelity}

According to the discussion of Section \ref{QT-section}, it turns
out that in order to quantify the efficiency of the present scheme
we need to calculate the two functions
\begin{eqnarray}\label{f-functions}
F_{\alpha_1}(x_1,x_2,\theta)&=&\bra{\alpha}_1 \rho_{1f} \ket{\alpha}_1, \\
F_{\alpha'_1}(x_1,x_2,\theta)&=&\bra{\alpha'}_1 \rho_{1f}
\ket{\alpha'}_1,
\end{eqnarray}
namely the fidelity with respect to states $\ket{\alpha}_1$ and
$\ket{\alpha'}_1$, respectively [as suggested by the notation such
functions do not depend on $\varphi$ due to Table 1 and Eq.
(\ref{rho''})]. Notice how the efficiency of the teleportation
scheme is maximum when either $F_{\alpha_1}(x_1,x_2,\theta)=1$ and
$F_{\alpha'_1}(x_1,x_2,\theta)=0$ or
$F_{\alpha_1}(x_1,x_2,\theta)=0$ and
$F_{\alpha'_1}(x_1,x_2,\theta)=1$. Using Eqs. (\ref{rho''}) and
(\ref{rho_1f}), such fidelities can be put in the form
\begin{eqnarray}\label{f-functions}
F_{\alpha_1}(x_1,x_2,\theta)&=&1-\frac{B\sin^2\frac{\theta}{2}}{A+B+C\cos\frac{\theta}{2}}, \\
F_{\alpha'_1}(x_1,x_2,\theta)&=&1-\frac{A\sin^2\frac{\theta}{2}}{A+B+C\cos\frac{\theta}{2}},
\end{eqnarray}
where $A$, $B$ and $C$ are functions of $x_1$ and $x_2$ according to
\begin{eqnarray}\label{f-functions}
A(x_1,x_2)&=&|\Phi_{0,1}^{+}(x_1)|^2|\Phi_{0,2}^{+}(x_2)|^2+|\Phi_{0,1}^{-}(x_1)|^2|\Phi_{0,2}^{-}(x_2)|^2,\label{A}\\
B(x_1,x_2)&=&|\Phi_{0,1}^{+}(x_1)|^2|\Phi_{0,2}^{-}(x_2)|^2+|\Phi_{0,1}^{-}(x_1)|^2|\Phi_{0,2}^{+}(x_2)|^2,\label{B}\\
C(x_1,x_2)&=&\left[|\Phi_{0,2}^{+}(x_2)|^2+|\Phi_{0,2}^{-}(x_2)|^2\right]\left[\Phi_{0,1}^{+}(x_1)\left(\Phi_{0,1}^{-}(x_1)\right)^*+\mathrm{c.}\,\mathrm{c.}\right],\label{C}
\end{eqnarray}
where the starred symbol stand for the complex conjugate. It turns
out that functions (\ref{A})-(\ref{C}) are real valued. Furthermore,
$A(x_1,x_2)$ and $B(x_1,x_2)$ can have only positive values. Using
these properties, the following inequalities hold
\begin{eqnarray}\label{f-functions}
F_{\alpha_1}(x_1,x_2,\theta)&\ge&F_{\alpha_1}=1-\frac{B}{A+B-|C|}, \\
F_{\alpha'_1}(x_1,x_2,\theta)&\ge&F_{\alpha'_1}=1-\frac{A}{A+B-|C|},  \\
\end{eqnarray}
Spatial functions $F_{\alpha_1}$ and $F_{\alpha'_1}$ therefore
represent $\theta$-independent lower bounds for
$F_{\alpha_1}(x_1,x_2,\theta)$ and $F_{\alpha'_1}(x_1,x_2,\theta)$,
respectively.
\begin{figure*}
\resizebox{0.95\columnwidth}{!}{\includegraphics{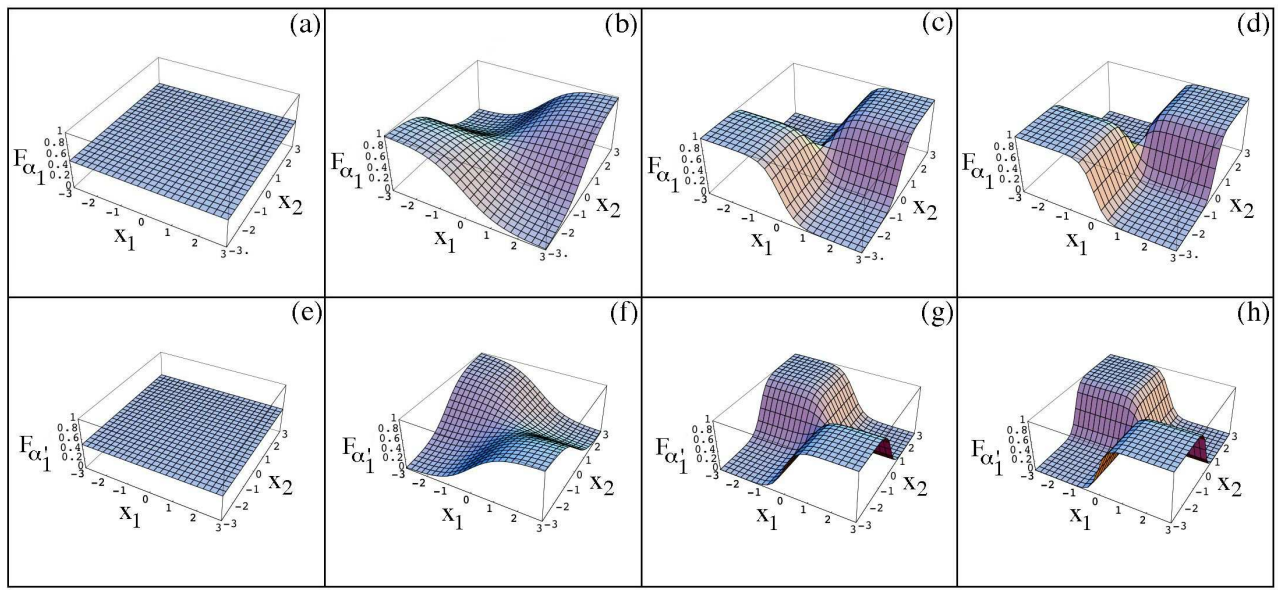}}
\caption{ Panels
(a), (b), (c) and (d): lower bound for the fidelity with respect to
$\ket{\alpha}_1$ $F_{\alpha_1}$ as a function of $x_1$ and $x_2$ at
the rescaled atom-cavity interaction times $\varepsilon\tau=$1 (a),
$\varepsilon\tau=$5 (b), $\varepsilon\tau=$8 (c) and
$\varepsilon\tau=$10 (d). Panels (e), (f), (g) and (h): lower bound
for the fidelity with respect to $\ket{\alpha'}_1$ $F_{\alpha'_1}$
as a function of $x_1$ and $x_2$ at the rescaled atom-cavity
interaction times $\varepsilon\tau=$1 (e), $\varepsilon\tau=$5 (f),
$\varepsilon\tau=$8 (g) and $\varepsilon\tau=$10 (h). The atomic
positions are in units of $\sigma_x$ (width of the initial atomic
wavepackets). The parameters used are: $\lambda=10^{-5}$m,
$\varepsilon = 10^5$ sec$^{-1}$, $m=10^{-26}$ kg and
$\sigma_x=\lambda/10$ \cite{nota-fig1}.} 
\label{Fig1}
\end{figure*}
For the sake of simplicity, we consider equal initial translational
states for both the atoms with zero mean position and  momentum
along the $x$-cavity axis. In such a case, the field-deflected
wavepackets $\Phi_{0,i}^+(x_i,\tau)$ and $\Phi_{0,i}^-(x_i,\tau)$
($i=1,2$) have the explicit form \cite{Vag-Cus}
\begin{equation}\label{deflected-wp}
\Phi_{0,i}^\pm(x_i,\tau)=\frac{e^{\mp\frac{i ma
\tau}{\hbar}x}e^{-\frac{\left(x\pm\frac{a
\tau^2}{2}\right)^2}{4\sigma_x^2+\frac{i\hbar
\tau}{2m}}}}{(2\pi)^{1/4}\sqrt{\sigma_x+\frac{i\hbar \tau}{2m
\sigma_x}}}
\end{equation}
where $a=(\hbar k\varepsilon/m)$. Notice that
$\Phi_{0,i}^+(x_i,0)=\Phi_{0,i}^-(x_i,0)$. Eq. (\ref{deflected-wp})
shows how the wavefunctions $\Phi_{0,i}^+(x_i,\tau)$ and
$\Phi_{0,i}^-(x_i,\tau)$ travel with constant acceleration $a$
towards the negative and positive semi-axis, respectively. It
follows that, provided the atom-field interaction time is large
enough (of the order of some $\varepsilon^{-1}$), the overlap
between the two packets becomes negligible and a measurement of the
atomic position is able to distinguish them. In Fig.~1 we plot the
lower bounds for the fidelity  $F_{\alpha_1}$ and $F_{\alpha'_1}$
against $x_1$ and $x_2$ at the rescaled atom-cavity interaction
times $\varepsilon\tau=1,5, 8,10$. Notice how at small times both
$F_{\alpha_1}$ and $F_{\alpha'_1}$ are non-zero and markedly lower
than 1. However, at larger times their maxima start approaching 1 up
to the point that, in the considered ranges of $x_1$ and $x_2$,
$F_{\alpha_1}=1$ in the domain $x_1x_2>0$ and $F_{\alpha_1}=0$ at
$x_1x_2<0$, whereas $F_{\alpha'_1}=1$ in the domain $x_1x_2<0$ and
$F_{\alpha_1}=0$ at $x_1x_2>0$. Such regime is reached provided
$\varepsilon\tau\simeq 10$ and it is therefore experimentally
accessible with present-day cavity coherence times
\cite{raymond,nota-regime}. In summary, these results provide clear
evidence that large enough atom-field interaction times allow to
perform efficient teleportation via atomic position measurements.

\section{Conclusions} \label{conclusions}

In summary, in this paper we have considered a cavity-QED
teleportation scheme that allows to teleport an unknown quantum
state between two atoms flying through a cavity via position
measurements on their translational degrees of freedom. Among the
major advantages of the scheme are the success probability of 1/2
and its ability to be performed both without direct projections onto
highly entangled subspaces and with no holonomous constraints on the
atom-cavity interaction times to be fulfilled. The detailed
measurement process able to give rise to successful teleportation
has been explicitly described in order to derive the final reduced
density matrix associated with the state of the target qubit. This
has allowed us to perform an efficiency analysis of the scheme in
terms of two fidelity functions. We have derived a lower bound for
each of them as a function of the possible outcomes of the position
measurements performed on the two atoms. Our analysis has shown that
a few Rabi oscillations are enough in order to attain the maximum
efficiency in a wide range of atomic positions.

Together with our previous study \cite{TC}, these results provide
strong evidence that the atomic translational dynamics in cavity QED
holds promises as an attractive resource to be harnessed in order to
perform quantum information processing tasks. Interestingly, such
idea supports a change of perspective given that the atomic
translational dynamics has been frequently addressed so far as an
unwanted effect that spoils quantum coherent phenomena in cavity-QED
systems \cite{Vag-Cus,which-path,epl-2atoms,epjd}. The use of such
degrees of freedom as a tool in order to either improve known
quantum information processing schemes, such as the generation of
maximally entangled states, or design novel ones is under ongoing
investigations.

\section*{Acknowledgements} Mauro Paternostro (Queen's University of
Belfast) is gratefully acknowledged for fruitful discussions. FC
acknowledges support from PRIN 2006 ``Quantum noise in mesoscopic
systems''.


\begin{thebibliography}{99}

\bibitem{nc} M. A. Nielsen and I. L. Chuang,  \textit{Quantum Computation and
Quantum Information} (Cambridge University Press, Cambridge, U. K.,
2000)
\bibitem{bennett} C. H. Bennett, G. Brassard, C. Crépeau, R. Jozsa, A. Peres, and W. K. Wootters, Phys. Rev. Lett. \textbf{70}, (1993) 1895
\bibitem{bowm-boschi} D. Bouwmeester, J.-W. Pan, K. Mattle, M. Eibl, H. Weinfurter, and A. Zeilinger,  Nature (London) \textbf{390}, (1997) 575
; D. Boschi, S. Branca, F. De Martini, L. Hardy, and S. Popescu,
Phys. Rev. Lett. \textbf{80}, (1998) 1121
\bibitem{NMR} M. A. Nielsen, E. Knill, and R. Laflamme, Nature (London) \textbf{396},
(1998) 52.
\bibitem{natphys}Q. Zhang, A. Goebel, C. Wagenknecht, Y.-A. Chen (Chen, Yu-Ao), B. Zhao, T. Yang T, A. Mair, J. Schmiedmayer, Nat. Phys. \textbf{2}, (2006) 678
\bibitem{nature} J. F. Sherson, H. Krauter, R. K. Olsson, B. Julsgaard, K. Hammerer, I. Cirac, E. S. Polzik, Nature (London) \textbf{443}, (2006) 557
\bibitem{raimond} J. M. Raimond, M. Brune, and S. Haroche, Rev. Mod. Phys. \textbf{73},
565 (2001)
\bibitem{teleportation-QED} L. Davidovich, N. Zagury, M. Brune, J. M. Raimond, and S. Haroche, Phys. Rev. A \textbf{50}, (1994) R895
; J. I. Cirac, and A. S. Parkins, Phys. Rev. A \textbf{50}, (1994)
R4441 ; S. B. Zheng and G. C. Guo, Phys. Lett. A \textbf{232},
(1997) 171 ; S. B. Zheng, Opt. Commun. \textbf{167}, (1999) 111 ; S.
Bose, P. L. Knight, M. B. Plenio, and V. Vedral, Phys. Rev. Lett.
\textbf{83}, (1999) 5158 ; S. Bandyopadhyay, Phys. Rev. A 62, (2000)
012308
\bibitem{vaidman} L. Vaidman, Phys. Rev. A \textbf{49}, (1994) 1473
\bibitem{de-almeida} N.G. de Almeida, R. Napolitano, and M. H. Y. Moussa, Phys. Rev A. \textbf{62}, (2000) 010101
\bibitem{zheng} S.-B- Zheng, Phys. Rev. A \textbf{69}, (2004) 064302
\bibitem{ye-guo} L. Ye, and G.-C. Guo, Phys. Rev. A \textbf{70}, (2004)
054303; R. W. Chhajlany and A. Wójcik, Phys. Rev. A \textbf{73},
(2006) 016302; L. Ye, and G.-C. Guo, Phys. Rev. A \textbf{73},
(2004) 016303
\bibitem{cardoso} W. B. Cardoso, A. T. Avelar, B. Baseia, and N. G. de Almeida Phys.
Rev. A 72, (2005) 045802
\bibitem{TC} M. Tumminello, and F. Ciccarello, to appear on Phys. Rev. A, arXiv:0706.0173v1 (2007)
\bibitem{Vag-Cus} A. Vaglica, Phys. Rev. A \textbf{58},
(1998) 3856; I. Cusumano, A. Vaglica, and G. Vetri, Phys. Rev. A
\textbf{66},  (2002) 043408
\bibitem{which-path} M. Tumminello, A. Vaglica, and G. Vetri, Europhys. Lett.
 \textbf{65},  (2004) 785
 \bibitem{SGE}T. Sleator, T. Pfau, V. Balykin, O. Carnal, and J.
Mlynek, Phys. Rev. Lett. \textbf{68}, (1992) 1996; C. Tanguy, S.
Reynaud, and C. Cohen-Tannoudji, J. Phys. B \textbf{17}, (1984) 4623
; M. Freyberger, and A. M. Herkommer, Phys. Rev. Lett. \textbf{72},
(1994) 1952 ; A. Vaglica, Phys. Rev. A \textbf{54},  (1996) 3195
\bibitem{schlicher} R. R. Schlicher, Opt. Comm. \textbf{70}, (1989) 97
\bibitem{wilkens} M. Wilkens, Z. Bialynicka-Birula, and P. Meystre, Phys. Rev. A \textbf{45}, (1992) 477
\bibitem{nota-nodale} For the benefit of the reader, in this paper we restrict to the nodal approximation of
the Hamiltonian since the mathematics in the antinodal case is a bit
more involved \cite{TC,epjd}. Nonetheless, we point out that both
the teleportation scheme \cite{TC} and the efficiency analysis
presented in this work hold in the antinodal case as well.
\bibitem{epl-2atoms} M. Tumminello, A. Vaglica, and G. Vetri, Europhys. Lett.
 \textbf{66},  (2004) 792
\bibitem{epjd} M. Tumminello, A. Vaglica, and G. Vetri, Eur. Phys. J. D
\textbf{36},  (2005) 235
\bibitem{vaglica95} A. Vaglica, Phys. Rev. A \textbf{52},
(1995) 2319
\bibitem{footnote}
Similarly to the seminal proposal by Bennett \emph{et al.}
\cite{bennett}, in such cases teleportation is completed after that
a 180 degree rotation around the $z$-axis in the internal Hilbert
space of atom 1 is performed.
\bibitem{nota-fig1} The considered range for each $x_i$ is such
that for $\varepsilon \tau<10$ the probability to find each atom in
the interval $[-3\sigma_x,3\sigma_x]$ is still significant.
\bibitem{raymond} S. Kuhr \emph{et al.}, Appl. Phys. Lett. \textbf{90},
(2007) 164101
\bibitem{nota-regime} Actually, this is an overestimated value since
a look at Fig.~1 shows how even at shorter strengths of
$\varepsilon\tau$, such as $\varepsilon\tau=5$, there is not so a
narrow range of position-measurement outcomes that ensure successful
teleportation.
\end{thebibliography}
\end{document}